\documentclass[journal]{IEEEtran}

\ifCLASSINFOpdf
\else
   \usepackage[dvips]{graphicx}
\fi
\usepackage{url}

\usepackage{graphicx,algorithm,algpseudocode,amsmath,bm,amsfonts,booktabs,subfigure}
\usepackage{cite}

\begin{document}

\title{Sequence-level Speaker Change Detection with Difference-based Continuous Integrate-and-fire}

\author{Zhiyun Fan${*}\thanks{* denotes equal contribution to this work.}$, Linhao Dong$*$, Meng Cai, Zejun Ma, and Bo Xu}

\maketitle

\begin{abstract}
Speaker change detection is an important task in multi-party interactions such as meetings and conversations. In this paper, we address the speaker change detection task from the perspective of sequence transduction. Specifically, we propose a novel encoder-decoder framework that directly converts the input feature sequence to the speaker identity sequence. The difference-based continuous integrate-and-fire mechanism is designed to support this framework. It detects speaker changes by integrating the speaker difference between the encoder outputs frame-by-frame and transfers encoder outputs to segment-level speaker embeddings according to the detected speaker changes. The whole framework is supervised by the speaker identity sequence, a weaker label than the precise speaker change points. The experiments on the AMI and DIHARD-I corpora show that our sequence-level method consistently outperforms a strong frame-level baseline that uses the precise speaker change labels.
\end{abstract}

\IEEEpeerreviewmaketitle
\vspace{-0.14cm}
\section{Introduction}
\label{sec:intro}
\IEEEPARstart{M}{ulti-party} interactions such as meetings and conversations are one of the most important scenarios for many speech and language applications \cite{sun2021combination}. Speaker change detection (SCD), the task of finding the time points that a new speaker starts to speak, is critical for such applications and has received increasing attention in recent years \cite{yin2017speaker,ge2017speaker,hruz2017convolutional,sari2020auxiliary}.

SCD is known as an important part of speaker diarization \cite{zhang2019fully, bredin2017pyannote}. It was previously modeled with distance-based methods \cite{ge2017speaker,anguera2012speaker,chen1998speaker}, which segment audio with a sliding window, and the distance of speaker embedding is used to decide whether a speaker change happens between the adjacent segments. Since the pitch varies saliently with speaker changes, some pitch-based methods detect speaker changes with the change in pitch \cite{yang2005pitch,abdolali2012novel,hogg2019speaker}.
More recently, there are some attempts at predicting the speaker change at the end of the neural network without relying on a distance metric \cite{yin2017speaker,hruz2017convolutional,yella2014artificial,sari2019pre}. 
Almost all these end-to-end systems are based on binary classification (i.e. change or not) to predict whether a speaker change happens between frames or segments.  
Since these methods rely on the precise speaker change labels, they are categorized into frame-level models.

Recently, the sequence-level modeling methods have made great progress in automatic speech recognition (ASR) \cite{graves2006connectionist,bahdanau2016end,vaswani2017attention,graves2012sequence}.
These models rely on different alignment mechanisms to conduct sequence transduction and have shown their performance advantages in comparison with the frame-level hybrid models \cite{chiu2018state,he2019streaming,gulati2020conformer}.
The success of the sequence-level model in ASR inspires us that it may be suitable for the SCD task.

In this paper, we model the SCD task from the perspective of sequence transduction. Specifically, we propose a novel encoder-decoder model to convert the input feature sequence to the speaker identity sequence. Inspired by the success of Continuous Integrate-and-fire (CIF) in the ASR field \cite{dong2020cif}, we design a difference-based continuous integrate-and-fire (DCIF) mechanism to bridge the encoder and decoder. The DCIF performs two functions in the SCD task, including 1) detecting the speaker changes and splitting the encoded sequence into segments according to the SCD results, 2) calculating segment-level speaker embeddings and firing them to the decoder. Then the decoder predicts the speaker identity. Based on the above framework, our method processes on the sequence level and removes the need for frame-level speaker change labels in the training. Besides providing the model framework, we also present several effective methods used to complete our sequence-level model, including 1) a DifferNet to estimate the speaker difference for the DCIF,  2) the length normalization to better represent the fired speaker embeddings, and 3) a multi-label focal loss (MLFL) to boost the training. 

We evaluate our sequence-level model on a real recording meeting corpus, AMI \cite{carletta2007unleashing} and DIHARD-\uppercase\expandafter{\romannumeral1} corpus \cite{ryant2018first}. After exploring three important model settings, our method achieves 86.76\% and 89.29\% harmonic mean (Hn) of purity and coverage on AMI and DIHARD-\uppercase\expandafter{\romannumeral1}, respectively, outperforming the Hn of 86.00\% and 88.09\% from a strong frame-level baseline \cite{bredin2017pyannote}. In addition, we provide the ablation study to evaluate the importance of the applied methods.
Our contributions are summarized as follows: 1) As far as we know, we are the first to address the SCD task as sequence transduction and propose a sequence-level SCD framework. 2) We design a DCIF mechanism to detect speaker changes and automatically calculate segment-level speaker embeddings according to the detected results. 3) We demonstrate that our sequence-level method achieves a better SCD performance with weaker supervision than a strong frame-level SCD baseline and release our code at https://github.com/zhiyunfan/SEQ-SCD. 

\begin{figure*}[ht]
\centerline{\includegraphics[width=0.8\linewidth]{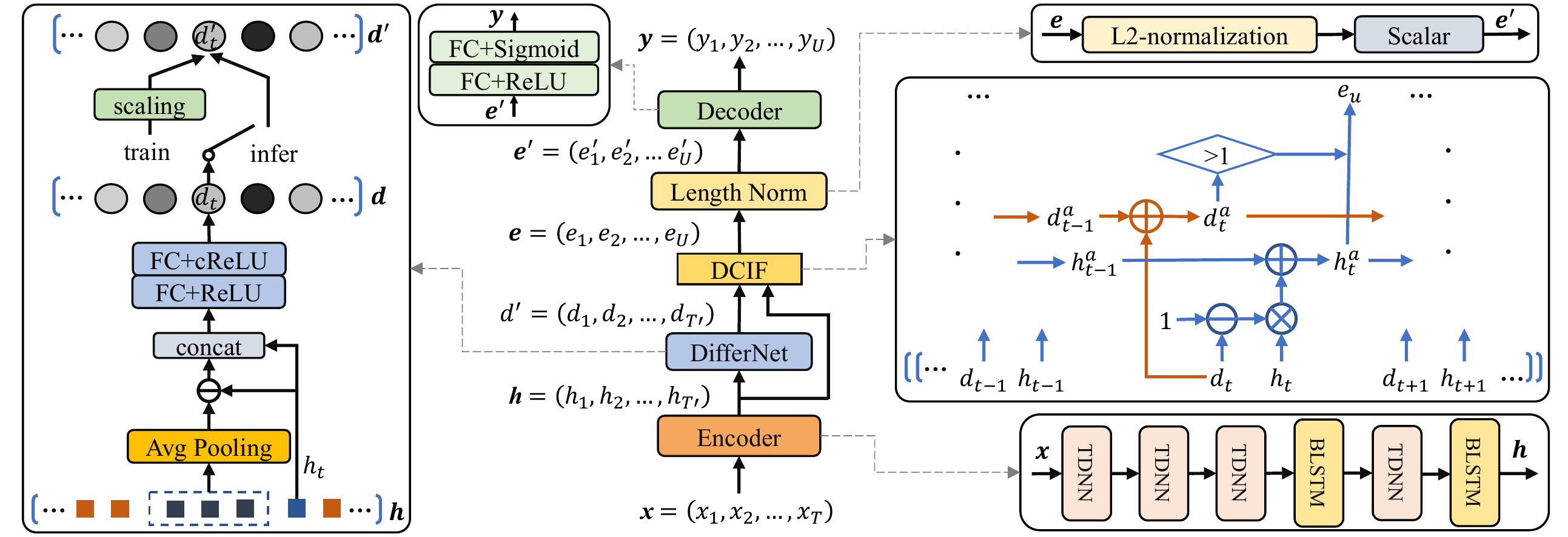}}
\setlength{\abovecaptionskip}{0.cm}
\setlength{\belowcaptionskip}{-0.6cm}
\caption{The architecture of sequence-level SCD model. The main body is in the middle of the diagram, and the details of the DifferNet, the DCIF, the length normalization layer and the encoder are shown in the boxes on both sides.}
\vspace{-0.5cm}
\label{model_fig}
\end{figure*}
\vspace{-0.14cm}
\section{Related work}
Most previous SCD methods \cite{yin2017speaker,ge2017speaker,anguera2012speaker,chen1998speaker,hruz2017convolutional,yella2014artificial,sari2019pre} detect speaker changes between frames or fixed-size windows by thresholding the distance or binary classification. These methods rely on the precise speaker change labels during training and are categorized into frame-level methods. In contrast, the sequence-level SCD first proposed in this paper addresses the SCD task as sequence transduction, which transfers the input feature sequence to the speaker identification sequence and predicts speaker changes by the specially designed DCIF. Benefit from the novel sequence-level model structure, our training process gets rid of the dependence on precise speaker change labels used in the frame-level methods. 

Our DCIF is inspired by the CIF in a sequence-level ASR model \cite{dong2020cif}. The CIF uses a pre-computed weight that scales the frame-level acoustic information contained in each frame to weight the frame-level representation and generates label-level representations. Different from the CIF, the DCIF integrates pre-computed speaker difference between each frame and its corresponding context instead of speaker information contained in each frame, and transfers frame-level speaker representations to segment-level speaker representations. Within the transfer, the weights of each frame and speaker difference are negatively correlated.


\vspace{-0.15cm}
\section{Speaker change detection as a sequence transduction task}
\vspace{-0.1cm}
\subsection{Sequence Transduction}
\label{ST}
We address SCD as a sequence transduction task. The input is a sequence of features, $x=(x_1,x_2,...,x_T)$, where $T$ is the total length of the input sequence. The output is a sequence of speaker identities $\boldsymbol y=(y_1,y_2,...,y_U)$, where $U$ is the number of segments of the input sequence after partitioning it on speaker change boundaries. Thus $U$$-$$1$ is the number of speaker change points. 
An encoder-decoder framework connected by a dynamic segmentation module is used for the sequence transduction of SCD.
The encoder transforms the input sequence to the frame-level speaker representations. The dynamic segmentation module detects speaker changes and splits the encoder outputs into segments according to the detected speaker change points. The speaker embeddings of the split segments are sent to the decoder for speaker classification. The objective is to find a function $f:x\rightarrow y$ that transforms the input feature sequence into a speaker identity sequence. The whole framework can be optimized as follow:
\begin{equation}
\setlength{\abovedisplayskip}{3pt} 
\setlength{\belowdisplayskip}{3pt}
    L=\frac{1}{U}\sum\nolimits_{u=1}^{U}\text{Classification\_loss}(f(x)_u,y_u)
\end{equation}
\vspace{-0.6cm}
\subsection{DCIF in the Sequence Transduction of SCD}
We propose a novel DCIF mechanism suitable for the SCD task to conduct the sequence transduction in Section \ref{ST}. It forwardly accumulates the speaker difference and integrates the speaker embedding simultaneously. Once the accumulated speaker difference reaches a threshold, the integrated speaker embedding will be fired for further speaker classification.
\renewcommand{\algorithmicrequire}{\textbf{Input:}} 
\renewcommand{\algorithmicensure}{\textbf{Output:}}
\begin{algorithm}[htb]
\caption{DCIF in the Sequence-level SCD.}
\label{alg:Framwork}
{\small{}
  \begin{algorithmic}[1]
    \Require
      The encoded frames $\boldsymbol h=(h_1,h_2,...,h_{T'})$, the speaker difference value $\boldsymbol d'=(d'_1,d'_2,...,d'_{T'})$, the threshold $\beta$;
    \Ensure
      The fired speaker embeddings $\boldsymbol e=(e_1,e_2,...,e_U)$;
    \State Initialize $u$$=$$1$, the accumulated speaker difference value $d_{0}^a$$=$$0$, the accumulated speaker embedding state $h_0^a=h_1$;
    \For{$t=1$; $t<=T'$; $t++$}
    \State // calculate currently accumulated speaker difference value
    \Statex \qquad \, and integrated speaker embedding state;
    \State $d_{t}^a = d_{t-1}^a + d'_t$;
    \State $h_t^a=h_{t-1}^a+(1-d'_t)*h_t$;
    \If{$d_t^a > \beta$}  // a speaker change is detected 
    \State // firing currently integrated speaker embedding state
    \State $e_u=h_t^a$; \, $u++$; 
    \State $h_t^a=h_t$;  // reset integrated speaker embedding state; 
    \State // $d'_t$ is divided into two part, the second part $d'_{t2}$ is
    \Statex \qquad \qquad used to reset accumulated speaker difference;
    \State $d'_{t1}=1-d_{t-1}^a$; \, $d'_{t2}=d'_t - d'_{t1}$;
    \State $d_t^a=d'_{t2}$;
    \EndIf
    \EndFor
    \State $e_U=h_t^a$; //save the speaker embedding for the last speaker;
    \label{code:fram:select} \\
    \Return $e=(e_1,e_2,...,e_U)$;
  \end{algorithmic}}
\end{algorithm}

The sequence-level SCD model is shown in Fig.\,\ref{model_fig}. The encoder and the decoder are connected by the DCIF. The DifferNet and the length normalization layer are designed to cooperate with the DCIF. Specifically, the encoder transforms the input feature sequence $\boldsymbol x=(x_1,x_2,...,x_T)$ to the frame-level speaker embedding $\boldsymbol h=(h_1,h_2,...,h_{T'})$. The DifferNet predicts the speaker difference $\boldsymbol d'=(d'_1,d'_2,...,d'_{T'})$, each of which is a scalar and corresponds to each encoded frame. 
The DCIF receives the frame-level speaker embedding $\boldsymbol h$ and the speaker difference $\boldsymbol d'$. Then it forwardly accumulates the speaker difference $d_t^a$ and integrates the speaker embedding $h_t^a$, which is an accumulation sum of $h_t$ weighted by $1-d_t'$. Once the accumulated speaker difference value reaches a threshold $\beta$, a speaker change point is located. Then the current speaker difference value will be divided into two parts, one for completing the current integration and the other for the next integration, and the currently integrated speaker embedding will be fired for further speaker classification. Until the last frame, we save the currently integrated speaker embedding for the last speaker. More details are shown in Algorithm \ref{alg:Framwork}.

In the inference stage, we save a mark sequence $c=[c_t \in \{0,1\}|t=1,...,T']$ along with the calculation of the DCIF. The $c_t=1$ indicates that the accumulated difference reaches the threshold at the $t$-th time step. The mark sequence is used for the calculation of SCD metrics.

\vspace{-0.14cm}
\section{Model Details}
\label{details}
To boost the performance of our sequence-level SCD, we propose the following three methods:

\textbf{DifferNet}: As shown in Fig.\,\ref{model_fig}, the DifferNet receives the frame-level speaker embedding $h_t$ and predicts speaker difference value $d'_t$ for each encoded frame. The speaker difference $d'_t$ is determined by $h_t$ and its corresponding history chunk. 
\begin{equation}
\setlength{\abovedisplayskip}{3pt} 
\setlength{\belowdisplayskip}{3pt}
    o_1=h_t-\frac{1}{l}\sum\nolimits_{\tau=t-l}^{t-1}h_\tau
\end{equation}
\begin{equation}
\setlength{\abovedisplayskip}{3pt} 
\setlength{\belowdisplayskip}{3pt}
   o_2=W_2*{\rm ReLU}(W_1*[o_1;h_t]+b_1)+b_2
\end{equation}
\begin{equation}
\setlength{\abovedisplayskip}{3pt} 
\setlength{\belowdisplayskip}{3pt}
\label{eq4}
    d_t=\text{min}(\text{max}(o_2, 0), 1)
\end{equation}
where $l$ is the length of the history chunk. $o_1$ and $h_t$ are concatenated and fed into the two FC layers. $h_t$ is used to reduce the interference of silence and noise in the measure of speaker difference $d'_t$. $W_1$, $W_2$, $b_1$, $b_2$ are trainable parameters. Equation \ref{eq4} is the formulation of cReLU \cite{choi2018pact} with an upper bound $1$. After finishing the calculation of speaker difference value $d_t$ for all encoded frames, the scaling operation is applied during training. 
\begin{equation}
\setlength{\abovedisplayskip}{3pt} 
\setlength{\belowdisplayskip}{3pt}
     d_t'= \kappa * d_t
\end{equation}
\begin{equation}
\setlength{\abovedisplayskip}{3pt} 
\setlength{\belowdisplayskip}{3pt}
    \kappa = (U-1) / \sum\nolimits_{t=1}^T{d_t}
\end{equation}
where $U$ is the length of the speaker identity sequence. The scaling operation ensures that the sum of $\boldsymbol d'$ is equal to the number of speaker changes $U$$-$$1$, which could make the number of fired speaker embeddings equal to the length of the speaker identity sequence. In the inference stage, the scaling operation is not used, which means that $d'_t$ is identical to $d_t$.

\textbf{Length Normalization}: The speaker embedding $e_u$ fired by the DCIF is a weighted sum of a varying number of frames ($e_u$ is detailed in Algorithm \ref{alg:Framwork}). We use an L2-normalization layer followed by a scalar \cite{cai2018analysis} to normalize the speaker embedding into a fixed hyperspace. The normalized embedding  $e_u'$ is euqal to $\eta * e_u / ||e_u||_2$.
$\eta$ is a hyper-parameter used to scale the unit-length speaker embedding into a fixed radius.
\textbf{Loss Function}: The loss function is the interpolation of a multi-label focal loss (MLFL) and a quantity loss \cite{dong2020cif}. 
\begin{equation}
\setlength{\abovedisplayskip}{3pt} 
\setlength{\belowdisplayskip}{3pt}
    L = \lambda_1 \frac{1}{U}\sum \limits_{u}{{\rm MLFL}(p_u, y_u)} + \lambda_2  \lvert U - 1 -\sum \limits_{t}{d_t'} \rvert
\label{LOSS}
\end{equation}

\begin{equation}
\setlength{\abovedisplayskip}{3pt} 
\setlength{\belowdisplayskip}{3pt}
\label{eq_mlfl}
\begin{split}
    {\rm MLFL}(p_u, y_u)=\frac{1}{C}\sum     \limits_{c}(-\alpha(1-p_{u,c})^\gamma y_{u,c}\text{log}(p_{u,c}) \\
     -(1-\alpha)p_{u,c}^{\gamma}(1-y_{u,c})\text{log}(1-p_{u,c}))
\end{split}
\end{equation}
where $y_u=[y_{u,c} \in \{0,1\} | c=1,...,C]$, and $y_{u,c}=1$ indicates that the speaker $c$ is presenting at segment $u$. The $p_u$ predicted by the decoder is the element-wise sigmoid activation for the $C$ speakers. MLFL is a combination of binary cross-entropy (BCE) \cite{fujita2019end} loss and focal loss \cite{lin2017focal}. Since $y_u$ may contain multiple speakers, we choose the BCE loss rather than the softmax. The focal loss makes the model focus on the positive samples and down-weight the numerous negative samples. The $\alpha$ and $\gamma$ are two hyper-parameters. 

The second item is a quantity loss which promotes the predicted firing times of the DCIF closer to the target number of speaker change points $U$$-$$1$. The $\lambda_1$ and $\lambda_2$ are two tunable hyper-parameters. 


\vspace{-0.2cm}
\section{Experiments and Results}
\subsection{Experimental Setup}
Experiments are performed on AMI \cite{carletta2007unleashing} and DIHARD-\uppercase\expandafter{\romannumeral1} corpus \cite{ryant2018first}. The AMI is a real-recorded $100$-hour English meeting corpus. We use Mix-Headset recordings for our experiments, and the division of the AMI corpus is consistent with the baseline system \cite{bredin2020pyannote}. 
For DIHARD-\uppercase\expandafter{\romannumeral1} corpus, we split the development set into two parts: $131$ files used as training set and the remaining $33$ files used as a new development set. The new development set is simply referred to development set in the following. We share the split at https://github.com/zhiyunfan/SEQ-SCD/tree/master/data/dihard1. 


For the model structure, the encoder stacks four Time Delay Neural Network (TDNN) layers and two Bi-LSTM layers. The details are shown in the lower right corner of Fig.\ref{model_fig}. The two Bi-LSTM layers both have 256 hidden units. The four TDNN layers have 512 channels with the context of [-2,-1,0,1,2], which sums up to five frames. The strides of the TDNN layers change with the number of temporal downsampling. (1,1,1,1), (1,1,1,2), (1,1,2,2), (1,2,2,2) and (2,2,2,2) is for $1/1$, $1/2$, $1/4$, $1/8$, $1/16$ downsampling, respectively. In the DifferNet, the length of the history chunk is explored in Section \ref{tuning}. The two FC layers are 512- and 1-dimensional, respectively. The hyper-parameter $\eta$ follows the best value $12$ in \cite{cai2018analysis}. For the DCIF, we set the $\beta$ to $1.0$. The decoder consists of two FC layers. The hidden layer has 256 units with ReLU activation. The output layer has 136 (the number of speakers in the training set) units with sigmoid activation. The loss hyper-parameters $\lambda_1$ and $\lambda_2$ is set to 50.0 and 1.0. The $\alpha$ and $\gamma$ in the MLFL are set to $0.25$ and $2$, the best value given by Lin \textit{et al}. \cite{lin2017focal}.

For the online processing, the input batch is randomly sampled from the raw session audio with a window. Then we apply additive noise from MUSAN dataset \cite{snyder2015musan} on-the-fly. The SNR values are sampled from $5$ to $20$ dBs. Specially, reverberation noise is used in Section \ref{comparison}. The room size is ranging from 2\,m-1\,m-2\,m to 10\,m-10\,m-5\,m (length-width-height). The wall absorption coefficient is sampled from 0.2 to 0.9. We extract 59-dimensional MFCC features (19 coefficients and energy with first- and second-order derivatives) with 25 ms frame length and $10$ ms frame shift. The batch size is $128$, and the length of the window is explored in Section \ref{tuning}. We use Adam \cite{kingma2014adam} optimizer, warming up the learning rate for the first $5\%$ of updates to a peak of $10^{-4}$, and holding on for the next $50\%$, and then linearly decaying for the remainder. 

During inference, the metrics of all models follow the tool of Pyannote \cite{bredin2017pyannote}. Firstly, we split each long test audio into fixed-length segments as same as training. And there is an $80\%$ overlap between two adjacent segments. For each frame, the final speaker change score is the average result of all segments containing this frame. Then the frames corresponding to prediction scores which are local maxima and greater than a tunable threshold $\theta$ are marked as speaker change points. All our experiments are evaluated on the purity, coverage \cite{bredin2017pyannote} and their harmonic mean (Hn). The tunable threshold is tuned on the development set to maximize the Hn.
\vspace{-0.3cm}
\subsection{Exploration on Model Settings}
\label{tuning}
Firstly, we explore three model settings in our sequence-level SCD model, including the size of the window used to sample batch, the temporal down-sampling in the encoder and the length of the history chunk used to calculate the speaker difference value. The size of the window affects the number of speaker change points in the batch. The temporal down-sampling decides the length of the encoded frame sequence fed into the DCIF. The length of the history chunk directly affects the calculation of speaker difference value.
\vspace{-0.2cm}
\begin{table}[h]
\centering
\setlength{\abovecaptionskip}{0cm}
\vspace{-0.2cm}
\caption{Evaluation of the sequence-level SCD on the development set of AMI with various model settings.}
\setlength{\tabcolsep}{2.6mm}{
\begin{tabular}{c|lccc}
\toprule
 & &  Purity & Coverage & Hn    \\
\midrule
&1\,s & - & - &  - \\
&2\,s &  77.39 &  90.79 &   83.55 \\
Size of window &4\,s &  81.39 &  87.41 &   \textbf{84.29} \\ 
&6\,s & 78.18 &  87.03 &  82.37 \\ 
&8\,s & 75.31 &  88.91  &   81.55 \\
\midrule
&1/1 &  81.39 &  87.41 & 84.29 \\ 
&1/2  &80.77 &  88.04 &  84.25  \\ 
Down-sampling&1/4  & 82.65 & 87.17 &   84.85  \\ 
&1/8  & 82.69 & 87.91 &  \textbf{85.22} \\
&1/16 & 81.95 & 87.26 &   84.52  \\ 
\midrule
&80\,ms &  82.87 & 87.21  &  84.99  \\
&160\,ms &  82.69 & 87.91 & \textbf{85.22} \\
Length of history&240\,ms & 80.80 &  89.70 &   85.02\\
&320\,ms & 81.58& 88.69 &84.99 \\
&400\,ms &  80.43 &  90.03 & 84.96   \\
\bottomrule
\end{tabular}
\label{table2}}
\end{table}
\vspace{-0.15cm}

Table \ref{table2} shows the results of our sequence-level model with various settings on the development set of AMI. In the upper part, we investigate the size of the window used to sample the batch. We fix a $1/1$ temporal down-sampling of the encoder and 160\,ms history chunk. As can be seen, the $4$\,s window achieves the best Hn value. Compared with the $2$\,s window, the $4$\,s window provides more speaker changes during training. But the performance degrades when the size of the window keeps increasing. Then we try to reduce the length through the temporal down-sampling of the encoder with a $4$\,s window and $160$\,ms history chunk. The results of various temporal down-sampling are shown in the middle of Table \ref{table2}. As the temporal down-sampling increases, our model obtains further performance gains. The $1/8$ temporal down-sampling gets the best results. The performance degradation of the $1/16$ may be due to multiple speaker change points covered by one encoded frame (There are a large number of rapid speaker change points in the AMI corpus). Finally, we fix a $4$\,s window and $1/8$ temporal down-sampling to compare the various length of the history chunk. The results are shown in the bottom part of Table \ref{table2}. We find that the Hn value slightly fluctuates with the changing of the length of the history chunk, and the $160$\,ms history chunk achieves the best performance. In summary, for our sequence-level SCD model, $4$\,s window to sample batch, $1/8$ temporal down-sampling, and $160$\,ms history chunk are relatively better model settings, which will be used for the subsequent experiments.

\vspace{-0.2cm}
\subsection{Comparison with Baseline}
\label{comparison}
\vspace{-0.4cm}
\begin{table}[ht]
\centering
\setlength{\abovecaptionskip}{0cm}
\caption{Evaluation of the baseline model and our best model on the test set of AMI and DIHARD-\uppercase\expandafter{\romannumeral1} corpora.}
\setlength{\tabcolsep}{1.4mm}{
\begin{tabular}{lcccccc}
\toprule
& \multicolumn{3}{c}{AMI}  & \multicolumn{3}{c}{DIHARD-\uppercase\expandafter{\romannumeral1}} \\ 
\cline{2-7}
& Purity & Coverage & Hn   & Purity & Coverage & Hn \\
\midrule
Pyannote \cite{bredin2020pyannote} & 83.00  &  89.30   & 86.00 & 84.99 & 91.43 &  88.09 \\
Ours & \textbf{83.92} & \textbf{89.81}& \textbf{86.76} &  \textbf{86.24} & \textbf{92.56} &  \textbf{89.29}  \\ 
\bottomrule
\end{tabular}}
\label{table4}
\end{table}
\vspace{-0.2cm}
In this section, we compare our model with the baseline model on the AMI and DIHARD-\uppercase\expandafter{\romannumeral1} corpora. The baseline results are achieved in an open-source toolkit, Pyannote \cite{bredin2020pyannote}. It directly predicts frame-level SCD results and applies the same Bi-LSTM layer as our model to conduct binary sequence labeling. Considering that the baseline model was trained for $1000$ epochs, we increase the training epoch of our model from $160$ to $500$. The results in Table \ref{table4} show that our sequence-level model consistently outperforms the baseline model on the two corpora. In addition, we compare our method with the baseline system on AMI corpus adding reverberation noise. The baseline and our method achieve 83.36\%-$88.86$\%-$86.02$\% and 84.25\%-89.83\%-$87.01$\% (Purity-Coverage-Hn), respectively. 
It reflects the robustness of the proposed sequence-level SCD method.

\vspace{-0.2cm}
\subsection{Ablation Study}
\vspace{-0.1cm}
\begin{table}[ht]
\centering
\setlength{\abovecaptionskip}{0cm}
\vspace{-0.3cm}
\caption{Evaluation of the ablation study on the development and test set of AMI. Results on the development set use a small font.}
\setlength{\tabcolsep}{2.9mm}{
\begin{tabular}{lccc}
\toprule
& Purity & Coverage & Hn   \\
\midrule
Full model & 83.92 \tiny 82.65 & 89.81 \tiny 88.56& 86.76 \tiny 85.50 \\ 
\hspace{1em} w/o Length Norm & 81.66 \tiny 79.91 & 91.24 \tiny 90.24 & 86.18 \tiny 84.76 \\ 
\hspace{1em} w/o Scaling& 83.87 \tiny 82.66 & 88.33 \tiny 87.27 & 86.05 \tiny 84.90 \\ 
\hspace{1em} w/o Focal Loss & 82.41 \tiny 81.97 & 87.97 \tiny 86.29 & 85.10 \tiny 84.08 \\ 
\bottomrule
\end{tabular}}
\label{table3}
\end{table}
\vspace{-0.15cm}
In this section, we use the ablation study to evaluate the importance of different methods applied to the sequence-level SCD model. As shown in Table \ref{table3}, the first row is the results achieved by the full model. In the following three experiments, we ablate the length normalization, the scaling operation and the focal loss, respectively. The results indicate that all three methods provide improvements. Among them, ablating the focal loss causes the largest performance degradation, 
which indicates that the focal loss alleviates the imbalance of positive and negative samples as we expected.

\vspace{-0.1cm}
\section{Conclusion}
In this paper, we address the speaker change detection task from the perspective of sequence transduction and propose a sequence-level SCD model using difference-based continuous integrate-and-fire (DCIF). 
Evaluated on the AMI and DIHARD-\uppercase\expandafter{\romannumeral1} corpora, our proposed sequence-level model achieves 86.76\% and 89.29\% harmonic mean (Hn) of purity and coverage without using any precise frame-level speaker change label, and outperforms the 86.00\% and 88.09\% Hn from a strong frame-level baseline \cite{bredin2017pyannote}. It demonstrates the effectiveness of the sequence-level model in the SCD task.

\bibliographystyle{IEEEtran}
\bibliography{refs}

\end{document}